\magnification=1200
\overfullrule=0pt

\font\ninerm=cmr9
\font\blackboard=msbm10 
\font\bigbf=cmbx10 scaled \magstep 2
\font\blackboards=msbm7 \font\blackboardss=msbm5
\newfam\black \textfont\black=\blackboard
\scriptfont\black=\blackboards \scriptscriptfont\black=\blackboardss
\def\Bbb#1{{\fam\black\relax#1}}

\def\e{\mathop{\rm e}\nolimits}
\def\Ai{\mathop{\rm Ai}\nolimits}
\def\d{{\rm d}}
\def\j{{\rm j}}
\def\mi{{\rm i}}
\def\G{\mathop{\Gamma}\nolimits}
\def\Re{\mathop{\rm Re}\nolimits}
\def\e{\mathop{\rm e}\nolimits}
\def\Ai{\mathop{\rm Ai}\nolimits}
\def\Arg{\mathop{\rm Arg}\nolimits}
\def\defi{\buildrel{\rm def}\over =}

\def\section#1{\nobreak{\bigskip\noindent{\bf#1}\par\medskip}}
\def\subsection#1{{\medskip{\bf#1}}}
\def\rf#1\par{\smallbreak\vskip-\parskip\item{[\hskip1pt#1\hskip-2pt]}}

\centerline{{\bigbf Airy function}\footnote{$^{\dag}$}
{Corrections (July 2003) indicated by footnotes.} }
\smallskip
\centerline{{\bf (exact WKB results for potentials of odd degree)}}
\bigskip
\centerline{{\bf Andr\'e Voros}}
\centerline{CEA--Saclay, Service de Physique Th\'eorique}
\centerline{F--91191 GIF-SUR-YVETTE CEDEX (France)\/}
\centerline{(e-mail: voros@spht.saclay.cea.fr)}

\vskip 24pt
{\narrower\ninerm 
An exact WKB treatment of 1-d homogeneous Schr\"odinger operators 
(with the confining potentials $q^N$, $N$ even) is extended to odd degrees $N$.
The resulting formalism is first illustrated theoretically and numerically 
upon the spectrum of the cubic oscillator (potential $|q|^3$).
Concerning the linear potential ($N=1$), the theory exhibits a duality 
in which the Airy functions $\Ai$, $\Ai '$ become paired
with the spectral determinants of the quartic oscillator ($N=4$).
Classic identities for the Airy function, as well as some less familiar ones,
appear in this new perspective as special cases in a general setting.
\bigskip}

A number of quantum spectral properties have now become established 
(by means of WKB theory, asymptotic or exact) for homogeneous $q^N$ potentials 
on the real line having a discrete spectrum (confining case, i.e., $N$ even).
After recalling the needed results (Sec.~1), this work will mainly extend 
and apply the framework to odd powers $N$ (Sec.~2), then specially discuss
the picture resulting for the Airy function when $N=1$ (Sec.~3).

\section{1. Summary of previous results.}
(See [1] for a more detailed review.)

\subsection{Definitions and notations:}
the Schr\"odinger operator, with all coefficients straightforwardly scaled out,
reads as [2--6]
$$ \hat H \defi -{\d^2 \over \d q^2}+q^N, \qquad 
\hbox{for $N$ an even integer.} \eqno(1)$$
(Obvious dependences upon the parameter $N$ will be left implied.)

Over $L^2(\Bbb R)$, $\hat H$ is a strictly positive operator;
it has a purely discrete (nondegenerate) spectrum 
$\{\lambda_k\}_{ k=0,1,2,\ldots}$, and $\lambda_k\uparrow +\infty$
according to the asymptotic Bohr--Sommerfeld law
$$ b_0 \lambda_k^\mu = 2\pi (k+1/2) + o(1),
\qquad k \to \infty {\rm \ \ in\ }\Bbb N,
 \eqno(2)$$
which uses the classical action $b_0 \lambda^\mu$ 
(at the energy $\lambda$, for the Hamiltonian $p^2+q^N$), with
$$ \mu \defi {N+2 \over 2N}:\quad 
{\rm the\ }growth\ order,\quad 
{\rm and} \quad b_0 = {2\pi^{1/2}\over N}\Gamma\Bigl({1 \over N}\Bigr)\Big/
\Gamma\Bigl({3 \over 2}+{1 \over N}\Bigr).  \eqno(3)$$
Other dynamical constants describe the phases $\alpha,\omega$ such that
$\hat H_{|L^2(\omega \Bbb R)}$ is unitarily equivalent to 
$\alpha \hat H_{|L^2(\Bbb R)}$:
this happens for $\omega = \alpha ^{-1/2}$ and $\alpha$ in the cyclic group
$\{ \e^{\mi \ell \varphi} \mid \ell=0,1,\ldots, L-1\}$, with
$$\varphi = {4\pi \over N+2}: \quad {\rm the}\ symmetry\ angle, \qquad
L = {N \over 2}+1: \quad {\rm the}\ symmetry\ order. \eqno(4)$$

Since $\hat H$ commutes with the parity operator $\hat P$, it splits as
$\hat H^\pm \defi \hat H(1 \pm \hat P)$;
the eigenfunctions of positive and negative parity 
carry the even and odd labels $k$ respectively. 
{\it Spectral functions\/}, defined next, may also be split accordingly.

\subsection{The spectral zeta functions:} [4,5]
$$Z(s) \defi \sum_k \lambda_k^{-s} \qquad  ({\rm and} \quad Z^\pm (s) \defi 
\sum_{k\ {\scriptstyle{\rm even}\atop \scriptstyle{\rm odd}}} \lambda_k^{-s} )
\qquad \qquad (\Re s > \mu); \eqno(5)$$
they extend {\it meromorphically\/} to all of $\Bbb C$, 
with only simple poles located at (or among) $s=(1-2j)\mu,\ j=0,1,2,\ldots$.
For $\Re s < \mu$ the series (5) diverge 
and the definitions must be regularized, according to eq.(2); 
e.g., in each parity sector, 
$$ \eqalign{
{\sum_k} ^\ast \lambda_k^{-s}\ &\defi \lim_{K \to +\infty} \Bigl\{ \sum_ {k<K}
\lambda_k^{-s} \ -{b_0 \mu \over 4\pi}{\lambda_{K}^{-s+\mu} \over (-s+\mu)} 
+{1\over 2} \lambda_{K}^{-s} \Bigr\}
\qquad {\rm if\ } \Re s > - \mu, \cr
&\hbox{with both $k,K$ even for $Z^+$, odd for $Z^-$.} \cr} \eqno(6)$$
In particular, $Z^\pm(0) = \pm 1/4$ (using eq.(2)), and
$$ {Z^\pm}'(0) = \lim_{K \to +\infty} \Bigl\{ -\sum_ {k<K}
\log \lambda_{k} 
\ +{b_0 \over 4\pi}\lambda_{K}^\mu \Bigl( \log \lambda_{K}-{1 \over \mu}\Bigr)
  -{1\over 2}\log \lambda_{K} \Bigr\} \quad
{\rm for\ }k,K\ \scriptstyle{\scriptstyle{\rm even}\atop \scriptstyle{\rm odd}}
\ , \eqno(7)$$
which in turn generates the ({\it zeta-regularized\/}) {\it determinants\/}, as 
$\det \hat H^\pm \defi 
\exp - {Z^\pm}'(0)$.
 
\subsection{The spectral (or functional) determinants:} [5,7] \quad
$ D^\pm(\lambda) \defi \det (\hat H^\pm + \lambda) $:

\noindent these are {\it entire\/} functions, more explicitly given by
$$ \eqalignno{
D^\pm(\lambda) &\equiv \exp [- {Z^\pm}'(0)] 
\prod_{k\ {\scriptstyle{\rm even}\atop \scriptstyle{\rm odd}}}
(1+\lambda/\lambda_k) \qquad \ \ ({\rm for\ } \mu < 1, \ i.e.,\ N>2) &(8) \cr
&= \exp \Bigl[ -{Z^\pm}'(0) 
- \sum_{n=1}^\infty {Z^\pm (n) \over n}(-\lambda)^n \Bigr] \quad 
({\rm for\ } \mu \ne 1,\ i.e.,\ N \ne 2,{\rm \ and\ }|\lambda|<\lambda_0). ~~~~
&(9) \cr
}$$

The harmonic oscillator ($N=2$) has a special status in the family: 
it is {\it the\/} solvable case but also {\it the\/} ``confluent" case, 
i.e., the growth order becomes {\it integer\/} ($\mu=1$),
and this invalidates several formulae in their generic form given here;
in particular, $Z^\pm(1)$ are infinite and eqs.(8--9) diverge, 
a valid substitute specification being
$$ D^\pm(\lambda) = 
2^{\pm 1/2} \sqrt{2\pi}\ 2^{-\lambda/2} 
\Big / \G \Bigl({2 \mp 1+\lambda \over 4}\Bigr) 
\qquad {\rm for\ }N=2 . \eqno(10) $$
Hence $N=2$, potentially the only elementary example for the formalism 
under review, is a pathological case instead.

The determinants admit semiclassical asymptotic expansions 
for $|\lambda| \to \infty$ ($ {\vert \arg \lambda \vert <\pi -\delta} $).
These have quite stringent forms, actually simpler for the full determinant
$D=D^+D^-$ and the ``skew" ratio $D^{\rm P} \defi D^+/D^-$ : [4]
$$ \eqalign{
\log D(\lambda) &\sim \sum_{j=0}^{\infty} a_j \lambda^{\mu(1-2j)} 
\quad (N \ne 2) \qquad 
({\rm with\ } a_0 \equiv (2 \sin \pi\mu)^{-1} b_0),  \cr
\log D^{\rm P}(\lambda) &\sim {1 \over 2} \log \lambda 
+ \sum_{r=1}^{\infty} d_r \lambda^{-(1+N/2)r}.  \cr} \eqno(11)$$
The leading behaviors imply that $D$ and $D^\pm$, as entire functions, 
are of order $\mu$
(the higher coefficients $a_j,\ d_r$ are computable term by term, too).

\subsection{Exact functional relations:} [8,1]
the determinants satisfy a basic functional relation, obtained by an exact WKB
calculation
(and corresponding to the reflection formula for $\G(z)$ when $N=2$):
$$ \eqalign{
\e^{\mi \varphi/4} D^+ (\lambda) D^-(\e^{\mi \varphi} \lambda)
-\e^{-\mi \varphi/4} D^+ (\e^{\mi \varphi} \lambda) D^- (\lambda) 
&\equiv 2 \mi, \qquad (N \ne 2) \cr
i.e., \qquad
\e^{\mi \varphi/4} D_0^+ D_1^- - \e^{-\mi\varphi/4} D_1^+ D_0^- 
&\equiv 2\mi, \cr}
 \eqno(12)$$
upon introducing the generic shorthand notation
$D_\ell(\cdot) \defi D(\e^{\mi \ell\varphi}\cdot)$
($\varphi$ being the symmetry angle, and $\ell$ an integer mod~$L$).
Within identities like (12), all such subscripts are globally shiftable mod~$L$,
since $\lambda$ can be freely rotated by $\e^{\mi\varphi}$ throughout.

Eq.(12) was found [1] as an equivalent form of a multiplicative 
coboundary formula linking the full and skew determinants, [5]
$$ \eqalignno{
 \e^{-\mi \varphi /2} 
{D^{\rm P}(\e^{\mi\varphi} \lambda) \over D^{\rm P}(\lambda)} 
&\equiv \exp [ -2 \,\mi \Phi(\lambda)], \qquad {\rm for\ }
\Phi(\lambda) \defi {\rm arcsin} \,\bigl[ 
\bigl( D(\e^{\mi\varphi}\lambda)D(\lambda) \bigr)^{-1/2} \bigr] , &(13) \cr
i.e.,\qquad {D_{\ell+1}^{\rm P} \over D_\ell^{\rm P} } &\equiv 
\exp {\mi( -2 \Phi_\ell + \varphi /2)},
\qquad \ell=0,1,\ldots,L-1 \ ({\rm mod~}L) \quad (N \ne 2).~~~~  &(14) \cr}$$
This in turn entails a consistency condition upon $D$, 
embodied in this {\it cocycle relation\/} (of length $L$), 
$$ \sum_{\ell=0}^{L-1} \Phi_\ell (\lambda) \equiv L \varphi/4 \quad 
= \pi/2 \quad {\rm for\ even\ } N \quad (N \ne 2). \eqno(15)$$
The full determinant $D$ thereby inherits an autonomous functional equation,
circularly symmetric of order $L$ (and convertible to a polynomial form).

\subsection{Spectral sum rules:} [5]
algebraic identities arise from expanding either functional relations (12) 
or (13) to all powers $\lambda^n$ around $\lambda=0$, with the help of eq.(9);
by eq.(5), the output is an infinite sequence of sum rules connecting various spectral moments:
$$ \eqalignno{
[Z'(0)=] \quad {Z^+}'(0) + {Z^-}'(0)
 & \equiv \log\,\sin \textstyle{\varphi \over 4}  &(n=0) \ (16.0) \cr
\sin \textstyle{\varphi \over 4}\ Z^+(1) 
- \sin \textstyle{3\varphi \over 4}\ Z^-(1) & \equiv 0  \qquad \qquad \qquad 
(N \ne 2) &(n=1) \ (16.1) \cr
\sin \textstyle{3\varphi \over 4}\ Z^+(2)
- \sin \textstyle{5\varphi \over 4}\ Z^-(2) & \equiv
\sin \textstyle{\varphi \over 4}
\, \bigl( 2 \cos \textstyle{\varphi \over 4}\ [Z^+(1)-Z^-(1)] \bigr)^2
 ~~~~~~~~~~~~~~~~~~~~~~ &(n=2) \ (16.2) \cr & \ \vdots \cr
\sin (n-\textstyle{1 \over 2}) \textstyle{\varphi \over 2}\ Z^+(n)
-\sin (n+\textstyle{1 \over 2}) \textstyle{\varphi \over 2}\ Z^-(n)
& \equiv  P_n\{Z^+(m),Z^-(m)\}_{1 \le m < n}  &({\rm rank\ } n)
\ (16.{\it n}) \cr}$$
where $P_n$ is a polynomial, homogeneous of degree $n$ under the ruling
that each $Z^\pm (m)$ is of degree $m$. The relative weights in the left-hand-sides recur with period $L$ in $n$, 
yielding $Z(n)$ whenever $n$ is a multiple of $L$. For such $n=0 \ {\rm mod~}L$,
$Z(n)$ can be further reduced to a polynomial in the $\{Z(m)\}_{1 \le m < n}$
only: this follows by directly expanding the closed functional equation 
for $D$ itself, eq.(15), in place of eq.(12)
(an option which however misses all the other identities of ranks 
$n \ne 0 \ {\rm mod~}L$).
By contrast, no $Z^+(n)$ or $Z^-(n)$ ever comes alone on a left-hand side, 
which precludes similarly closed functional equations 
for $D^+$ or $D^-$ separately.

\subsection{Exact quantization conditions:} [6,1]
exact quantization formulae {\it \`a la\/} Bohr--Sommerfeld for the spectrum
ultimately derive from eq.(12), as
$$\eqalign{
2\Sigma_+(\lambda_k) &= k+{1 \over 2}+{\kappa \over 2},\qquad k=0,2,4,\ldots \cr
2\Sigma_-(\lambda_k) &= k+{1 \over 2}-{\kappa \over 2},\qquad k=1,3,5,\ldots \cr
}
\qquad {\rm with\ } \kappa \defi {N-2 \over N+2} \ , \eqno(17)$$
where, excepting the case $N = 2$ (singular but solvable otherwise, 
to $ \Sigma_\pm(\lambda) \equiv \lambda/4$),
$$ \Sigma_\pm(\lambda) \defi \pi^{-1} \Arg D^\pm(-\e^{-\mi\varphi} \lambda)
\ ^{\dag} \qquad (\lambda \ge 0),
 \eqno(18')$$
$$ \Arg D^\pm(-\e^{-\mi\varphi} \lambda) =
\sum_{k'\ {\scriptstyle{\rm even}\atop \scriptstyle{\rm odd}}}\phi_{k'}(\lambda)
,\ ^{\dag} \qquad 
\phi_{k'}(\lambda) \defi \arg(\lambda_{k'} - \e^{-\mi\varphi} \lambda);
\eqno(18'')$$
i.e., $\Arg D^\pm$ denotes the determination of $\arg D^\pm$ which is continuous
over the half-line $[0,-\e^{-\mi\varphi}\infty)$ \footnote{$^{\dag}$}
{$-\e^{-\mi\varphi}$ and not $\e^{-\mi\varphi}$ as in the printed version.}
and vanishing at $\lambda=0$, 
and this is the sum of the angles $\phi_{k'}(\lambda)\ (\in [0,\pi)$) 
subtended by the vector $\overrightarrow{(0,\lambda)}$
at the points $\e^{\mi\varphi} \lambda_{k'}$.

The system (17--18) appears to specify the exact spectrum 
{\it selfconsistently\/} within each parity sector.
It amounts to a fixed-point condition for the (nonlinear) mapping defined by
the application of eqs.(18) followed by (17), upon sequences of trial levels
$\lambda _k^{(0)}$ that are {\it asymptotically correct\/} 
([6,1]: to be more precise, we require their compliance with eq.(2)).

This fixed-point mechanism is moreover easy to implement numerically
within some preset tolerance $\varepsilon$, which has to be nonzero to render 
the computational scheme finite-dimensional (by inducing a high-$k$ cutoff): 
upon such truncations, then, the straightforward loop iterations of these maps
appear to converge towards the correct spectra, 
consistently within $O(\varepsilon)$, and at geometric rates [6]. 
Such a {\it constructive\/} enactment of eqs.(17--18) works if,
essentially, some form of {\it contractivity\/} holds for the underlying map.
What we now have in this direction is no rigorous statement yet, 
but the numerical observation of such a behavior in every tried case,
plus partial analytical evidence: e.g., the mapping can be simplified
by treating the spectrum as continuous [1], in which approximation 
its contraction factor is just $\kappa$ --- and $|\kappa|<1$ by eq.(17).

\section{2. Extension to odd degrees and applications.} 
The above framework will now be strengthened in two ways, both involving
a violation of analyticity through the invocation of {\it odd\/} degrees $N$, 
in which case the confining potential becomes $|q|^N$.

\subsection{From even to odd $N$:}
all previous results actually follow from solving 
a connection problem for the Schr\"odinger equation 
$ (\hat H + \lambda) \psi(q) = 0 $ along a {\it half-line\/}, 
e.g., from $q=0$ to $+\infty$. 
The required {\it complex WKB\/} calculations were performed {\it exactly\/} [5]
thanks mostly to the analyticity of the potential 
(whose homogeneity also eased the operations but at a more concrete level).

Consequently, these procedures apply to {\it odd\/} polynomials $q^N$ as well, 
with the understanding that all results refer to the spectra specified by 
the same boundary conditions {\it on the half-line\/} $[0,+\infty)$ 
as in the case of even $N$: specifically, at $q=0$, Neumann for positive parity,
resp. Dirichlet for negative parity.
But these in turn precisely select the positive, resp. negative parity spectra 
of the potential $|q|^N$ {\it also when $N$ is odd\/}.

{\it All\/} previous formulae extend to odd $N$ under this interpretation, 
with just one forced explicit change:
the order $L$ of the symmetry group, being an integer, cannot retain the value
$N/2+1$ from eq.(4), but becomes twice that instead:
$$L=N+2  \quad {\rm for\ odd\ } N , \eqno(19)$$
while $\varphi$ {\it stays unchanged\/}; 
consequently, the cocycle relation (15) becomes
$$\sum_{\ell=0}^{L-1} \Phi_\ell (\lambda) \equiv L \varphi/4 \quad = \pi \quad 
{\rm for\ odd\ } N. \eqno(20)$$

\smallskip
\noindent $\bullet$ Example: $N=3$.
We may test the extended formalism upon the confining homogeneous 
{\it cubic oscillator\/} (the operator $\hat H = -{\d^2 / \d q^2}+|q|^3$), 
for which
$$ \mu=5/6, \quad \varphi = 4\pi/5, \quad L=5; \quad 
b_0={2^{2/3} \sqrt 3 \over 5\pi} \G(1/3)^3 = a_0;
\quad \kappa=1/5.  \eqno(21)$$

Among the several results which can be validated,
we describe the exact quantization formulae (17--18)
as they probe the functional relation (12) the most sharply.
So, we iterated eqs.(18--17) (under a tolerance value $\varepsilon =10^{-9}$)
upon the input data consisting of each 
parity subsequence of the semiclassical spectrum $\lambda _k^{(0)}$ 
(given by $b_0\, \lambda _k^{(0)\ 5/6} = 2\pi (k+{1 \over 2})$), 
and we readily observed convergence to these lowest eigenvalues:
$$ \matrix{
k & {\rm even} & ~~~~~~~~~ & k & {\rm odd} \cr
0 & 1.0229479 & & 1 & 3.4505627 \cr
2 & 6.3702932 & & 3 & 9.5220764 \cr
4 & 12.870297 & & 5 & 16.369373 \cr
6 & 20.000879 & & 7 & 23.745471 \cr
8 & 27.592421 & & 9 & 31.530790 \cr
}\eqno(22)$$
The convergence appears to be governed by contraction factors
$\approx +0.233$ for positive parity ($k,\ k'$ even)
and $+0.189$ for negative parity ($k,\ k'$ odd) 
(cf. the continuous-spectrum prediction $=\kappa=1/5$).

Separately, we diagonalized the matrix of $\hat H$
in the orthonormal eigenfunction basis $\{\psi_n\}$ 
of the harmonic oscillator $-{\d^2 / \d q^2}+q^2$, using
$$ \langle \psi_{n'} |\ |q|^N \ | \psi_{n''} \rangle =
\sqrt{2^{n'+n''}n'! \ n''! \over \pi} 
\sum_{{\scriptstyle{0 \le m' \le n'/2}\atop \scriptstyle{0\le m'' \le n''/2}}}
\Bigl( -{1 \over 4} \Bigr)^{m'+m''}
{\G \bigl({1 \over 2}(1+N+n'+n'') -(m'+m'')\bigr) 
\over m'! \ m''! \ (n'-2m')! \ (n''-2m'')!} 
\eqno(23)$$
for $n'= n''$ mod~2, zero otherwise; 
in each parity sector we applied several truncations (in the size range 20--40),
and retained the eigenvalue figures as far as they were fully stable. 
These results then showed complete agreement with eq.(22). 
(Here, the brute-force calculation might well be the less reliable one: 
eq.(23) generates numerous large entries, 
through cancellations between even larger terms.)

The exact method is thus validated to an accuracy of $10^{-6}$--$10^{-7}$,
whereas its input (we used the semiclassical spectrum) was off by as much as
$10^{-1}$ (for the ground state: $\lambda _0^{(0)} \approx 0.920791$).

\smallskip
\noindent $\bullet$ General $N$: an alternative exact approach exists 
for odd and even degrees alike:
it takes the fully analytic potential $q^N$ (nonconfining for $N$ odd) 
and privileges a different spectral function, 
namely a {\it Stokes multiplier\/} --- normalized here as 
$C(\lambda) \equiv \e^{-\mi\varphi /4} c(\lambda)$ relative to refs.~[9,2].
The present analysis connects to that approach and recovers its results, 
through the identity
$$ C_0 \equiv (2\mi)^{-1} 
(\e^{\mi \varphi/2} D_0^+ D_2^- - \e^{-\mi\varphi/2} D_2^+ D_0^-)
\qquad ({\rm for\ } N \ne 2). \eqno (24)$$

For instance, when $N=3$, $C(\lambda)$ obeys a closed functional equation, 
$ C_2 C_3 - C_0 \equiv 1 $
(the subscripts being shiftable mod~5; cf. [9]; [2] chap.5, sec.27; [10]).
Here, this equation is not just verified by the expression (24) (using eq.(12)),
but even {\it solved\/} thereby, to the extent that 
$D^\pm$ can be constructed by using the fixed-point eqs.(17--18) 
to get their zeros (cf. eq.(22)) and then forming the Hadamard products (8).
(By contrast, we do not know how to specify 
any of the other entire functions, like $D$ or $C$, as directly).

(Note: remarkably, the same $N=3$ functional-equation structure 
$C_2 C_3 - C_0 \equiv 1$ also appears in integrable 2D field-theory models 
that involve dilogarithms [11], and in the I-st Painlev\'e function [12].)

\subsection{Duality:}
within homogeneous problems, it is worthwhile to ask
which analytic potentials $q^N,\ q^{N'}$ might share 
the same rotation symmetry group, since the latter imprints the structure 
of spectral functions in a basic way. The answer is a {\it duality\/} relation,
$$ \eqalign{ 
\varphi + \varphi' &= 2\pi  \cr 
{\rm implying} \qquad {1 \over \mu} + {1 \over \mu'} =2, \qquad L &= L' , 
\qquad \kappa' = - \kappa,  \qquad {\rm and} \quad NN'=4 . \cr
} \eqno(25)$$
One solution is $N=N'=2$: thus the harmonic oscillator is selfdual  
(but also singular). However, there exists one (and only one) other solution:
the pair $N=4$ (homogeneous quartic oscillator [5]) 
and $N'=1$ (Airy equation [13]).

Hence, the Airy function (as the relevant solution of the linear potential)
turns out to be conjugated with the spectral determinants of $q^4$~!
This form of duality may be weak
(not implying exact links between the solutions of both problems) 
but still liable to reflect strong structural resemblances
(now that the framework accommodates even and odd $N$ on the same footing).

\section{3. Airy function vs quartic oscillator.}
We will now basically scan properties of the Airy function ($N=1$) [13]
as they appear in the global setting of the $|q|^N$ problem, 
and especially analyze its duality with the quartic case ($N=4$) [4--5].

Already, (unlike $N=2$,) both $N=1$ and $N=4$ share the status of 
{\it regular\/} values for the whole formalism
as they give {\it non-integer\/} growth orders $\mu$. 
Thus, the various constants corresponding to $N=4$ are
$$ \mu=3/4, \quad \varphi = 2\pi/3, \quad L=3; \quad 
b_0={\sqrt{2/\pi} \over 3} \G(1/4)^2, \quad a_0= b_0 / \sqrt 2;
\quad \kappa=1/3.  \eqno(26)$$
Then (almost) all previous formulae also hold in their 
{\it raw (generic) form\/} for $N=1$, upon simply resetting all the constants:
$$ \mu=3/2, \quad \varphi = 4\pi/3, \quad L=3; \quad b_0=8/3, \quad a_0=-4/3;
\quad \kappa=-1/3.  \eqno(27)$$

\subsection{Special features of the Airy case:} 
when $N=1$, the eigenvalues are those of the potential $|q|$, 
namely (up to sign) the {\it zeros\/} of the Airy function for negative parity 
and those of its derivative for positive parity
(the connection with their usual notation [13] being 
$a_s \equiv -\lambda_{2s-1}$,
\quad $a'_s \equiv -\lambda_{2s-2}$,\quad  $s=1,2,\ldots$);
the Bohr--Sommerfeld rule (2) (plus its corrections, omitted here) reproduces
their known asymptotic behavior.
The corresponding spectral determinants are given by
$$ D^-(\lambda)=2 \sqrt \pi \Ai(\lambda), \qquad 
D^+(\lambda)=-2 \sqrt \pi \Ai'(\lambda) .  \eqno(28)$$
(Normalization, the only non-obvious issue, is determined by adjusting the
asymptotic forms (11), which are expressly without constant terms,
against the known asymptotic expansions of $\Ai$ and $\Ai'$ [13].)

Thus, the extension of the formalism to odd $N$ reveals its {\it only\/}
elementary regular example, given by the Airy functions
(notwithstanding that their underlying potential $|q|$ is the most singular one
at 0).

Still, one amendment is specially mandated when $N=1$, 
namely a (standard) regularization [7]
for those series and products which turn divergent as $\mu \ge 1$: 
thus, $Z^\pm(1)$ are now computed from eq.(6) instead of (5), 
and eq.(8) for $D^\pm$ accordingly gets replaced by
$$ D^\pm(\lambda) = \e^{-{Z^\pm}'(0)+Z^\pm(1) \lambda} 
\prod_{k\ {\scriptstyle{\rm even}\atop \scriptstyle{\rm odd}}}
(1+\lambda/\lambda_k)\e^{-\lambda/\lambda_k}. \eqno(29)$$

\subsection{The basic determinantal identities:}
we first compare the functional relations (12) and their consequences 
for the two conjugate cases, 

$$\eqalign{
\e^{\mi\pi/6} D^+(\lambda)D^-(\j\lambda)
-\e^{-\mi\pi/6} D^+(\j\lambda)D^-(\lambda) & \equiv 2\mi \qquad 
{\rm for\ }N=4 \quad (\varphi=2\pi/3) \cr
\e^{\mi\pi/3} D^+(\lambda)D^-(\j^2\lambda)
-\e^{-\mi\pi/3} D^+(\j^2\lambda)D^-(\lambda)& \equiv 2\mi \qquad 
{\rm for\ }N=1 \quad (\varphi=4\pi/3) \cr
} \qquad (\j \defi \e^{2\mi\pi/3}).  \eqno(30)$$
The latter, by eq.(28), is simply the classic Wronskian relation between
$\Ai(\cdot)$ and $\Ai(\j^2 \cdot)$, which both solve the same Airy equation:
$${\rm W}[\Ai(\cdot),\Ai(\j^2\cdot)] \equiv (2\pi)^{-1}\e^{\mi\pi/6}.
\eqno(31)$$
Here the Airy function $\Ai(z)$ acts in two ways, 
as solution {\it and\/} as spectral determinant of the Airy equation; 
this confusion of roles stems from the property (specific to $N=1$) that 
the operator $\hat H + \lambda$ solely involves the combination variable 
$z=q + \lambda$.

For $N>1$, by contrast, the determinants are not known to solve any second-order
differential equation; while this remains to be confirmed in full generality,
the fact is already certain for $N=2$
(due to the presence of $\G(z)$: cf. eq.(10), and [2], chap.5, sec.27).
Still, the $N=1$ and $N=4$ functional relations are highly similar
(duality!); but their different phase prefactors will suffice to
create an essential distinction (as eq.(33) will show).

If the discrete symmetry rotations are performed upon either of eqs.(30), 
a closed system of $L=3$ equations follows:
$$á\eqalign{
\e^{\mi \varphi/4} D_1^+ D_2^- - \e^{-\mi\varphi/4} D_2^+ D_1^- &\equiv 2\mi \cr
\e^{\mi \varphi/4} D_2^+ D_0^- - \e^{-\mi\varphi/4} D_0^+ D_2^- &\equiv 2\mi \cr
\e^{\mi \varphi/4} D_0^+ D_1^- - \e^{-\mi\varphi/4} D_1^+ D_0^- &\equiv 2\mi .
\cr } \eqno(32)$$

With the $D_\ell^-$ as unknowns, this is a linear system, 
whose $3 \times 3$ determinant has the value 
$$\Delta = 2\mi \,\sin 3\varphi/4 \ D_0^+ D_1^+ D_2^+ . \eqno(33)$$

\smallskip
\noindent $\bullet \ N=4$: $\varphi=2\pi/3$, thus $\Delta \not\equiv 0$; 
then the odd spectral determinant $ D^- (= D_0^-)$ can be solved 
as a functional of the even one (and vice-versa), in rational terms:
$$ D_0^- \equiv {D_0^+ -\j^2 D_1^+ -\j D_2^+ \over D_1^+ D_2^+}\ ,\qquad 
{\rm and} \qquad D_0^+ \equiv {D_0^- -\j D_1^- -\j^2 D_2^- \over D_1^- D_2^-} 
\eqno(34)$$
(the subscripts being shiftable mod~3).

\smallskip
\noindent $\bullet \ N=1$: $\varphi=4\pi/3$, thus $\Delta \equiv 0$, 
a sharply different situation. Now the system cannot be solved for the $D_\ell^-$, 
and the elimination process yields a (linear) relation among these instead,
(and likewise for $D^+$),
$$ \eqalign{
D_0^- + \j^2 D_1^- +\j^4 D_2^- \equiv 0, \quad i.e., \quad
\Ai(\cdot) + \j \Ai(\j \,\cdot)  + \j^2 \Ai(\j^2 \,\cdot) &\equiv 0 , \cr
D_0^+ + \j D_1^+ + \j^2 D_2^+ \equiv 0, \quad i.e., \quad
[\Ai(\cdot) + \j \Ai(\j \,\cdot)  + \j^2 \Ai(\j^2 \,\cdot)]' &\equiv 0 ; \cr}
\eqno(35)$$
hence both identities simply reflect the classic 3-solution dependence relation 
for the Airy equation (but now in partnership with the nonlinear eqs.(34)).

\subsection{The cocycle functional equations:}
their comparison will already uncover a less obvious (new?) identity 
for the Airy functions.

\smallskip
\noindent $\bullet \ N=4$: in the quartic case, the cocycle relation (15)
(of length 3) implies $\sin \Phi_2=\cos(\Phi_0 +\Phi_1)$; 
expanding and squaring the $\cos \Phi_\ell$ away in succession yields 
$2 \sin \Phi_0 \sin \Phi_1 \sin \Phi_2 
 + \sin^2 \Phi_0 + \sin ^2\Phi_1 + \sin^2 \Phi_2 =1$ and then, 
by the definition of $\Phi$ in eq.(13), [5]
$$ D_0 D_1 D_2 \equiv D_0 + D_1 + D_2 +2 . \eqno(36)$$

\smallskip
\noindent $\bullet \ N=1$: the search for a non-trivial Airy counterpart 
of eq.(36) must also be directed at the full spectral determinant, 
which now reads as
$$ D = -4 \pi \Ai \Ai' = -2 \pi (\Ai^2)' \ ; \eqno(37)$$
the relevant cocycle relation has again length $L=3$,
but this time it is eq.(20); it now implies $\cos \Phi_2=-\cos(\Phi_0 +\Phi_1)$ 
and then, proceeding just as before,
$$ \eqalign{
D_0^2 + D_1^2 + D_2^2 - 2(D_1D_2 + D_2D_0 + D_0D_1) +4 &=0  \cr
D_0 =-2\pi(\Ai^2)'(\cdot),\qquad D_1 =-2\pi(\Ai^2)'(\j^2\,\cdot),\qquad 
D_2 &=-2\pi(\Ai^2)'(\j\,\cdot). \cr} \eqno(38)$$
Thus, $(\Ai^2)'$ exhibits a functional equation which is 
nonlinear, inhomogeneous, and with ternary symmetry like eq.(36) 
(but rederivable by elementary means, {\it a posteriori\/}).

\subsection{The Stokes multipliers:}
here, their expression (24) plus the functional relations (12), (36)
lead to diverging behaviors in the two cases.

\smallskip
\noindent $\bullet \ N=4$: the Stokes multiplier $C(\lambda)$ and
the full determinant $D(\lambda)$ get related both ways,
$$D_0 \equiv C_1 C_0 -1 \qquad \Longleftrightarrow
\qquad  C_0 \equiv (D_0 D_2-1)^{1/2} \eqno(39) $$
(the subscripts being shiftable mod~3);
hence $C$ stands equivalent to $D$ as a spectral function,
and indeed it has a functional equation very close to eq.(36), [10]
$$ C_0 C_1 C_2 \equiv C_0 + C_1 + C_2. \eqno(40) $$

\smallskip
\noindent $\bullet \ N=1$: $C(\lambda)$ degenerates to a constant
($\equiv 1$) [2] hence it has lost all spectral information.
(More generally, $D(\lambda)$ might be unrelated to $C(\lambda)$ for odd $N$.)

Therefore, duality is devoid of content under this specific angle.

\subsection{The spectral sum rules:}

\noindent $\bullet \ N=4$: for the quartic case, 
the simple substitution $\varphi=2\pi/3$ in eqs.(16) yields [5]:
$$ \eqalignno{ Z'(0) &= - \log 2 &(41.0) \cr
{1\over 2} Z^+(1) - Z^-(1) &= 0 &(41.1) \cr
Z^+(2) - {1\over 2} Z^-(2) &= {3 \over 2} (Z^+(1)-Z^-(1))^2 &(41.2) \cr
Z(3) &= {1 \over 6} Z(1)^3-{1 \over 2} Z(1)Z(2) \qquad {\rm etc.} &(41.3) \cr}$$
The left-hand-side coefficients then recur with period $L=3$ in $n$;
all $Z(n)$ with $n=3p$ reduce to polynomials in the $\{Z(m)\}_{1 \le m < n}$
only, deducible directly from the functional equation (36) for $D$ alone.

\smallskip
\noindent $\bullet \ N=1$: by setting $\varphi=4\pi/3$ in eqs.(16), 
similar sum rules are produced for the {\it moments of the zeros\/}
of $\Ai(\lambda)$ for odd parity, resp. $\Ai'(\lambda)$ for even parity
(i.e., $Z^-(n) = \sum_s (-a_s)^{-n}$, $Z^+(n) = \sum_s (-a'_s)^{-n}$),
$$ \eqalignno{ 
Z'(0) &= - \log (2/ \sqrt 3) \quad [={Z^+}'(0) + {Z^-}'(0)] &(42.0) \cr
Z^+(1) &= 0 &(42.1) \cr
Z^-(2) &= Z^-(1)^2 &(42.2) \cr
Z(3) & = Z^-(1)^3 - {3 \over 2} Z^-(1)Z^+(2) 
= {5 \over 2} Z(1)^3 - {3 \over 2} Z(1)Z(2) \qquad {\rm etc.} &(42.3) \cr}$$
All properties of the quartic case persist ($L=3$ again) but in addition here,
the closure property of identities within zeta values of the same parity label 
holds not only for $Z(3p)$ but also for $Z^+(3p+1)$ and $Z^-(3p+2)$, 
thanks to the autonomous functional equations (35) for $D^+$ and $D^-$
respectively.

Now, moreover, the Airy function is also a special function, 
whose Taylor series is accessible by other means [13], 
and this amounts to a full knowledge of the expansion (9). 
The ``spectral" sum rules above
thereby get completed by the ``special" sum rules below, in which
$\rho \defi -\Ai'(0)/\Ai(0) = 3^{5/6}(2\pi)^{-1} \G(2/3)^2 \approx 0.729011133$:
$$ \eqalignno{ 
{Z^+}'(0) - {Z^-}'(0) &= -\log \rho \qquad {\rm (by\ definition)} &(43.0) \cr
Z^-(1) &= -\rho   &(43.1) \cr
Z^+(2) &= 1/\rho   &(43.2) \cr
Z^-(3) = -\rho ^3+1/2 \qquad &(\Longleftrightarrow) \qquad Z^+(3) = 1 
\qquad {\rm (etc.)} &(43.3) \cr}$$
This last identity is exceptional for being so simple 
(the {\it inverse cubes of the zeros of\/} $\Ai'(-z)$ sum up to {\it unity\/}),
and {\it rational\/}: all higher $Z^\pm (n)$ can be thus expressed when $N=1$,
but as {\it nontrivial\/} (rational) functions of $\rho$,
hence are {\it never rational-valued\/} again.

Eqs.(42--43) remind of known sum rules for integer powers of zeros 
of the Bessel functions $J_\nu$ ([14], and refs. therein) 
but do not strictly correspond to them,
since $\Ai(z)$ is a Bessel function of a {\it non-integer\/} power of $z$
(and we know of no earlier explicit mention of any sum rule for Airy zeros).
Both sets of identities however have the same abstract basis: 
they express the consistency between two full specifications 
of the given function, a Taylor series like eq.(9)
and a Hadamard product like eqs.(8),(29).

\smallskip
\noindent $\bullet$ For general $N\ge 2$, by contrast, the $N=1$ explicit values
have extensions {\it of the same form\/} only for ${Z^\pm}'(0)$ and $Z^\pm (1)$ 
([4], [5] App. C--D: these refs.
stated the results for $N=2M$ only [4], but this limitation is unwarranted):
$$ \eqalign{
{Z^-}'(0) &= 
\log \Bigl[ (N+2)^{N \varphi \over 8\pi} \sqrt \pi 
\Big/ \G \Bigl({\varphi \over 4 \pi} \Bigr) \Bigr] \qquad \qquad \qquad \qquad
\Bigl( \varphi = {4 \pi \over N+2} \Bigr) \cr
Z^+(1) - Z^-(1) &= 
(2\sqrt\pi)^{-1} \Bigl[{2 \over N+2} \Bigr]^{N \varphi \over 2\pi}
\sin {\varphi \over 4} 
\ \G \Bigl({\varphi \over 4 \pi} \Bigr) \G \Bigl({2\varphi \over 4 \pi} \Bigr)
\G \Bigl({3\varphi \over 4 \pi} \Bigr) \Big/
\G \Bigl({1 \over 2} + {\varphi \over 2\pi} \Bigr)
 \cr} \eqno(44)$$ 
(the values for other parity labels follow through the identities (16.0--1)).

\subsection{The exact quantization of eigenvalues:}
the exact quantization mechanism (17--18) is mainly governed 
by the symmetry angle $\varphi$, and 
it degenerates, with quantization becoming fully explicit, 
for $\varphi=\pi$ (harmonic case). 
With respect to this ``critical" system, 
the quartic oscillator spectrum (with $\varphi=2\pi/3,\ \kappa=+1/3$) 
and the Airy zeros (with $\varphi=4\pi/3,\ \kappa=-1/3$) 
assume exactly mirror-symmetric positions; 
in particular, a key ingredient of the exact formula, the kernel function
$\arg(\lambda' - \e^{-\mi\varphi} \lambda)$
(associated with the linear ``flux operator" [6,1]),
takes opposite values in the two cases.

However, this symmetry is lost at some stage of the procedure
(and we see no manifest link between the two resulting spectra either). 
Thus, the system (17--18) is consistent and stable under iteration
only when applied to spectra having the right growth order $\mu$,
which differs in the two cases (3/4 for $N=4$, vs. 3/2 for $N=1$).
These consequences follow:

\smallskip
\noindent $\bullet \ N=4$: since $\mu <1$, eqs.(17--18) hold in their original
form (fig.1, upper part).
When applied to (definite-parity) input sequences $\lambda _k^{(0)}$
only subject to eq.(2) 
(i.e., $b_0\ \lambda _k^{(0)\ 3/4} \sim 2\pi (k+{1 \over 2})$, 
with $b_0$ from eq.(26)), 
the iteration scheme exhibits contractive convergence 
to the exact eigenvalues, with ratios
$\approx +0.392$ for positive parity ($k,\ k'$ even)
and $+0.333$ for negative parity ($k,\ k'$ odd) [6]. 

\smallskip
\noindent $\bullet \ N=1$: now $\mu >1$ hence the series in (18$''$) diverges
like $(\lambda \sin \varphi)$ times the series (5) for $Z^\pm (1)$; 
its proper regularization, as dictated by eq.(6) for $s=1$ and eq.(29), 
is then \footnote{$^{\dag}$}{See footnote at eqs.(18).}
$$ \Arg D^\pm(-\e^{-\mi\varphi} \lambda)\ =
\lim_{K \to +\infty}
\Bigl\{ \sum_{k'<K}  \phi_{k'}(\lambda)
- \lambda  \sin \varphi \,{2 \over \pi}\, \lambda _{K}^{1/2} \Bigr\} \qquad 
{\rm for\ }k',K\ \scriptstyle{\scriptstyle{\rm even}\atop \scriptstyle{\rm odd}}
\ . \eqno (45)$$
Here the summands $\phi_{k'}(\lambda)$, still defined as in eq.(18$''$) 
but with $\varphi=4\pi/3$, are all {\it negative and decreasing\/}
(fig.1, lower part); 
it is now the counterterm $=+\lambda (\sqrt 3 /\pi) \lambda _{K}^{1/2}$
which outweighs the whole sum to produce a positive and increasing 
left-hand side, as required for eq.(17).

Thereupon, the numerical scheme behaves for $N=1$ as for the quartic case. 
When applied to each parity subsequence of the semiclassical spectrum 
$\lambda _k^{(0)}$ 
(given by ${8 \over 3} \lambda _k^{(0)\ 3/2} = 2\pi (k+ {1 \over 2})$), 
the iteration of eqs.(17),(18$'$),(45) exhibits convergence to the Airy zeros,
with (roughly estimated) contraction factors 
 $-0.37$ for positive parity (zeros of $\Ai';\ k,\ k'$ even)
and $-0.25$ for negative parity (zeros of $\Ai;\ k,\ k'$ odd). 
(However, whereas for $N>2$ this scheme is comparatively quite  
efficient, for $N=1$ it should not beat special-purpose algorithms for 
obtaining the Airy zeros; also in this case, the regularization (45) proceeds
through large cancellations which strongly degrade the final accuracy.)

In conclusion, exact fixed-point quantization works for 
the spectra of Airy zeros too. Then, moreover, the conjugation symmetry shows
this spectrum quantization to be {\it exactly\/} as nontrivial 
as that of the quartic oscillator, even though the Airy equation itself 
{\it is\/} more elementary than its quartic 
--- or even its harmonic (and trivially quantized) --- analog.

\subsection{Acknowledgments}

We are very grateful to J. Robbins (Bristol, GB) for his seminal suggestion
of a possible link of the functional relation (12) with Wronskians 
(private discussion); 
to F.W.J. Olver (Maryland, USA) for thorough bibliographical guidance 
on Airy functions;
and to R. Tateo (Saclay and Amsterdam) and Y. Takei (RIMS, Kyoto) for acquainting us with refs. [11] and [12] respectively.

\bigskip
\centerline{\bf References}
\medskip
 
\rf 1

A. Voros, in: {\it Quasiclassical Methods\/} (IMA Proceedings, 
Minneapolis 1995), J. Rauch and B. Simon eds., IMA series vol. {\bf 95},
Springer, New York (1997) 189--224.

\rf 2

Y. Sibuya, {\it Global Theory of a Second Order Linear Ordinary 
Differential Operator with a Polynomial Coefficient\/}, 
North-Holland, Amsterdam (1975). 

\rf 3
 
C.M. Bender, K. Olaussen and P.S. Wang, Phys. Rev. {\bf D16} (1977) 1740--1748.
 
\rf 4
 
A. Voros, in: {\it The Riemann Problem, Complete 
Integrability and Arithmetic Applications\/}, eds. D. Chudnovsky 
and G. Chudnovsky, Lecture Notes in Mathematics {\bf 925}, Springer, Berlin 
(1982) 184--208 [augmented version of: Nucl. Phys. {\bf B165} (1980) 209--236].
 
\rf 5

A. Voros, Ann. Inst. H. Poincar\'e {\bf A 39} (1983) 211--338.

\rf 6

A. Voros, J. Phys. {\bf A 27} (1994) 4653--4661.

\rf 7

A. Voros, Commun. Math. Phys. {\bf 110} (1987) 439--465.

\rf 8

A. Voros, in: {\it Zeta Functions in Geometry\/}
(Proceedings, Tokyo 1990), eds. N. Kurokawa and T. Sunada, Advanced Studies in
Pure Mathematics {\bf 21}, Math. Soc. Japan, Kinokuniya, Tokyo (1992), 327--358.

\rf 9

Y. Sibuya and R.H. Cameron, in: 
{\it Symposium on Ordinary Differential Equations\/} 
(Proceedings, Minneapolis 1972), W.A. Harris Jr and Y. Sibuya eds.,
Lecture Notes in Mathematics {\bf 312}, Springer, Berlin (1973) 194--202.

\rf 10

Y. Sibuya, {\it On the functional equation
$f(\lambda)+f(\omega\lambda)f(\omega^{-1}\lambda) = 1,\ (\omega^5=1)$\/}, in:
R.C.P.~25 (Proceedings, 38$^{\rm e}$ Rencontre entre Physiciens Th\'eoriciens et
Math\'ematiciens, June 1984) vol. {\bf 34}, IRMA, Strasbourg (1984) 91--103.

\rf 11

Al.B. Zamolodchikov, Phys. Lett. {\bf B253} (1991) 391--394;
F. Gliozzi and R. Tateo, Int. J. Mod. Phys. {\bf A11} (1996) 4051--4064 
and refs. therein.

\rf 12

A.A. Kapaev, Differential Equations {\bf 24} (1989) 1107--1115 
[translated from the Russian (1988)].

\rf 13

M. Abramowitz and I.A. Stegun, {\it Handbook of Mathematical Functions\/} 
(Chap. 10.4), Dover, New York (1965); F.W.J. Olver, {\it Asymptotics and
Special Functions\/} (Chap. 11), Academic Press, New York (1974), and:
{\it Airy Functions\/}, Chapter 11, Digital Library of Mathematical
Functions Project, National Institute of Standards and Technology,
Gaithersburg, MD 20899-8910, U.S.A. (http://math.nist.gov/DigitalMathLib/);
O. Vall\'ee and M. Soares, {\it Les fonctions d'Airy pour la physique\/},
Diderot \'Editeur, Paris (1998).

\rf 14

G.N. Watson, {\it Theory of Bessel Functions\/} (chap. 15.41), Cambridge Univ.
Press (1952); A. Erd\'elyi (ed.), {\it Higher Transcendental Functions\/}
(Bateman Manuscript Project) (vol.~II, chap. 7.9), McGraw-Hill, New York (1953).

\vfill\eject


\topinsert
\vskip 13cm
$$\vbox to 12cm{\hbox to 15cm{\includegraphics{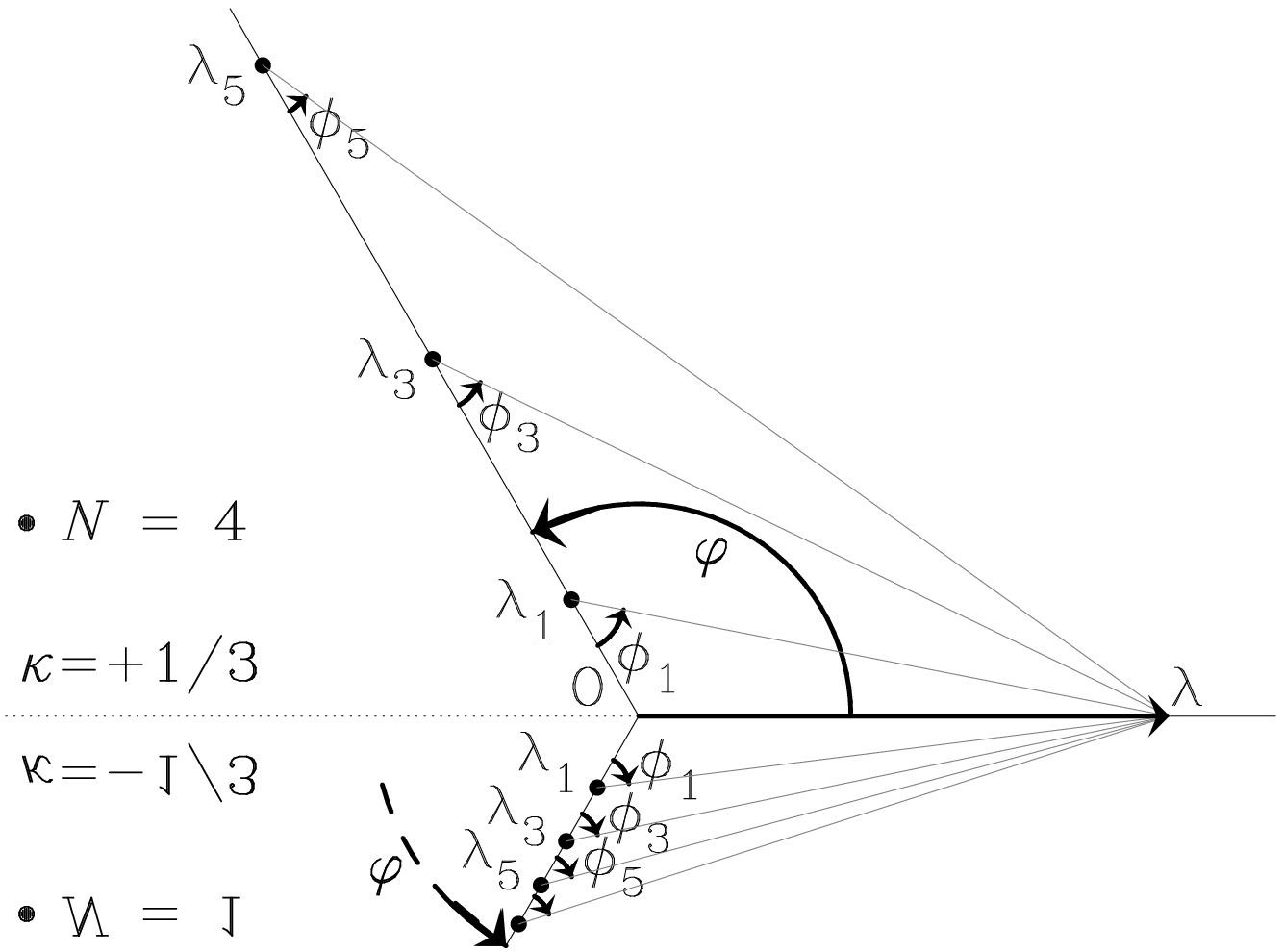}\hfill}\vfill}$$
\vskip -11cm
\noindent Fig.~1. Geometrical depiction of the exact-quantization summands 
$\phi_{k'} (\lambda)$ (at $\lambda = 15.$), in the odd spectral sector.
Upper half-plane: for the quartic oscillator, 
using $\varphi=2\pi/3$ in eq.(18$''$);
lower half-plane: for the zeros of $\Ai(-\lambda)$, 
using $\varphi=4\pi/3$ in eq.(45).
\endinsert

\end